❏ 1

# Comparative Simulation of Phishing Attacks on a Critical Information Infrastructure Organization: An Empirical Study

**Patsita Sirawongphatsara[1], Phisit Pornpongtechavanich[2], Nattapong Phanthuna[3], Therdpong Daengsi[4]**

[1]Computer Science Department, Faculty of Science and Technology, Rajamangala University of Technology Tawan-ok, Chonburi, Thailand

[2]Information Technology Department, Faculty of Industry and Technology, Rajamangala University of Technology Rattanakosin, Prachuap Khiri Khan, Thailand

[3]Electrical Engineering Department, Faculty of Engineering, Rajamangala University of Technology Phra Nakhon, Thailand

[4]Sustainable Industrial Management Engineering Department, Faculty of Engineering, Rajamangala University of Technology Phra Nakhon, Bangkok, Thailand

| Article Info | ABSTRACT |
|---|---|
|  | Nowadays, cybersecurity is crucial. Therefore, cybersecurity awareness should be a concern for businesses, particularly critical infrastructure organizations. The results of this study, using simulated phishing attacks, indicate that in the first attempt, workers of a Thai railway firm received a phony email purporting to inform recipients of a special deal from a reputable retailer of IT equipment. The findings showed that 10.9% of the 735 workers fell for the scam. This demonstrates a good level of awareness regarding cyber dangers. The workers who were duped by the initial attack received awareness training. Next, a second attempt was carried out. This time, the strategy was for the workers to change their passwords through an email notification from the fake IT staff. According to the findings, 1.4% of the workers fell victim to both attacks (different email content), and a further 8.0% of the workers who did not fall victim to the first attack were deceived. Furthermore, after the statistical analysis, it was confirmed that there is a difference in the relationship between the workers and the two phishing attack simulations using different content. As a result, this study has demonstrated that different types of content can affect levels of awareness. |

*Corresponding Author:*

Therdpong Daengsi
Department of Sustainable Industrial Management Engineering, Faculty of Engineering,
Rajamangala University of Technology Phra Nakhon
1381 Pibul Songkhram Road, Bangsue, Bangkok, Thailand
E-mail: therdpong.d@rmutp.ac.th

## 1. INTRODUCTION

### 1.1. Background and Significance

Cybersecurity has become a crucial global agenda; therefore, many countries have established organizations or agencies to address cybersecurity and related issues, such as the United Kingdom (UK) and Malaysia. In Thailand, the National Cyber Security Agency (NCSA) was officially established in 2019 to tackle cybersecurity-related issues [1]. The key objectives of this organization include developing policies, measures, and guidelines to maintain cybersecurity in both public and private organizations involved in critical information infrastructure. The importance of this endeavor lies in its capacity to prevent, respond to, and mitigate cyber threats, ensuring the security and stability of the state and domestic tranquility. The National Cybersecurity Committee (NCSC) is designated as the responsible agency by law, tasked with coordinating efforts between the public and private sectors. This collaboration is essential for the effective prevention and response to severe cybersecurity threats, ultimately fostering a secure and resilient cyber environment.

As noted in [2], the NCSA (National Cybersecurity Agency) of Thailand issued the Cybersecurity Act in recent years. This act outlines cybersecurity policies and action plans. Empowered by Section 49 of the Cybersecurity Act, the NCSA is authorized to develop general cybersecurity policies and action plans. Additionally, this agency has the authority to establish minimum requirements for computer systems and related hardware used by government organizations and Critical Information Infrastructure (CII) entities [1]. CII entities encompass banking and financial firms that store customer financial and sensitive information, as well as healthcare organizations that store patient information and medical records [3]. Moreover, transportation and logistics fall within the scope of CII entities, with electric railway systems categorized under transportation and logistics. Consequently, all businesses operating electric trains are required to comply with the Cybersecurity Act and oversee cybersecurity policies and strategies. Furthermore, all CII firms must enhance the cybersecurity proficiency of their staff to mitigate potential disruptions. Such incidents could significantly impact individuals who rely on electric trains for transportation.

One of the railway system companies in Thailand, operating in Bangkok and other metropolitan areas, initiated a program with the primary goal of enhancing cybersecurity awareness among its staff and ensuring compliance with the Cybersecurity Act. An academic institution collaborated with the train company on this initiative to assess the cybersecurity awareness levels among the company's employees.

One of the railway system companies in Thailand, operating in Bangkok and other metropolitan areas, initiated a campaign aimed at enhancing cybersecurity awareness among its staff and ensuring compliance with the Cybersecurity Act. Collaborating closely, an academic institution and the railway company jointly conducted an assessment to measure the level of cybersecurity awareness among the company's employees. This assessment constituted the main objective or research question of the study. The investigation led to the discovery of several intriguing cybersecurity problems. Presenting and discussing these concerns with other businesses interested in raising the cybersecurity awareness of their staff members, or with organizations that are comparable and categorized as Critical Information Infrastructure (CII) organizations, is vital. Furthermore, academics in the discipline may find these problems to be useful case studies or instances. This could be considered one of the major contributions or highlights of this article.

This article follows an earlier article [2] and consists of four sections. The first section serves as an introduction, providing an overview of cybersecurity awareness, phishing, and prior related works. The methodology is demonstrated in Section 2, followed by the presentation of results, analysis, and discussion in Section 3. Lastly, the conclusion is presented in Section 4.

## 1.2. Overview of Cybersecurity Awareness

Cybersecurity can be defined as a set of best practices, principles, policies, safeguards, guidelines, assurance, actions, risk management techniques, training, tools, and technologies that can be applied to secure users' assets and the surroundings of the business [4]. In the cyber environment, these assets may include infrastructure, networked hardware, services, applications, telecommunications systems, and various types of information [4].

"Awareness" is used throughout this text to describe cybersecurity awareness (CSA) or information security awareness (ISA). It is the understanding of cybersecurity and how to react appropriately to threats or attacks from the internet [5]. In order to combat cybersecurity threats, it is imperative that users and staff receive cybersecurity awareness training. This training can be delivered in a variety of ways, increasing user awareness and encouraging secure behavior with the ultimate goal of enabling users to identify and report malicious activity promptly and to continue using best cybersecurity practices [6]. The organization's cybersecurity awareness program and the goal to increase staff cybersecurity knowledge should be put into action together. With customized software, users or staff in each company can receive training, education, and increased awareness on how to protect their companies and themselves from cyberattacks (like phishing).

As depicted in Figure 1 and elaborated below, cybersecurity relies on three key factors: people, process, and technology [2]. The human element poses the highest risk of error due to its inherent unpredictability compared to software systems. Therefore, the workforce requires adequate resources, awareness, and training to serve as the primary defense against cyberattacks. Process encompasses the deployment of suitable technology and employee training. Management system approaches, risk assessments, frameworks, and audits are among the processes that can be implemented. However, the effectiveness of procedures and processes depends on the competence of the individuals implementing them. Hence, knowledgeable staff members are crucial.Technology plays a critical role in controlling and mitigating the risk of cyberattacks, especially in sectors like the public sector where numerous systems and individuals share records and sensitive data. Given the reliance on data access, it underscores the importance of employing appropriate software to safeguard these processes and data access points. Establishing a robust cybersecurity foundation is essential to minimize the impact of cyberattacks. Each organization must successfully integrate people, processes, and technology to develop a resilient defense against cyber threats.

### 1.4. Phishing

Cyber criminals typically target workers in every business in an effort to get access to corporate networks and computer systems in order to steal user and customer data. Phishing is a tactic employed by cyber criminals that poses a significant risk to Internet users. As mentioned in [7], phishing attack may come in a form of phishing email that will redirect users to the infected website and ask them to provide their sensitive information (e.g., personal information and bank account number), which will be used by the attacker to steal important information or money in the bank account. Phishing attack is usually difficult to identify, while the attacker can disguise the URL of the phishing website to work like the legitimate one [7]. Phishing uses social engineering techniques in addition to technological ones (see Figure 2 [5]). Nevertheless, the general design of the phishing assault is shown in Figure 3 [5].

The phishing-related facts in last few years are fascinating and can be stated as follows:
•   The most common kind of cybercrime reported to the US Internet Crime Complaint Center was phishing, which impacted around 300,000 persons [8].

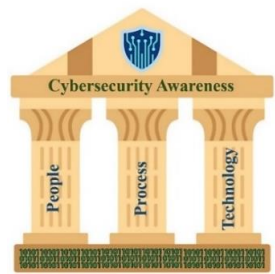

Figure 1. Three cybersecurity awareness pillars

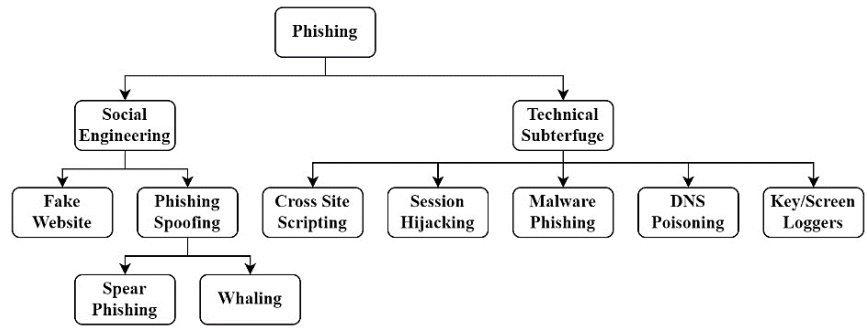

Figure 2. Phishing techniques

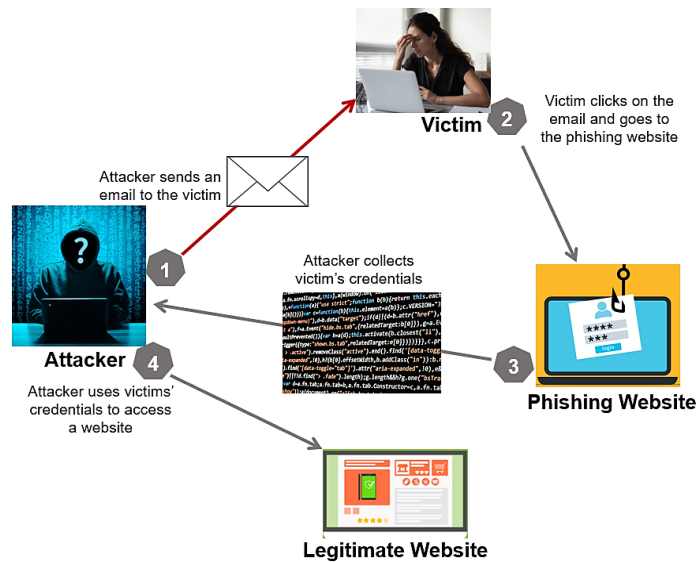

Figure 3. Overview on phishing techniques

• Firms reported breaches involving client or customer data, at the rate of 54% in 2021 and 44% in 2022. In 2022, 43% of the participants stated that they were often emailed ransomware infections [9].

• The main targets of phishing attacks were delivery companies, which received more than 27% of all global financial phishing attacks. E-stores ranked second, accounting for over 16% of the attacks, while payment systems ranked third with 10.39% [10].

Attackers are now employing more advanced techniques to increase the likelihood of successfully breaching information systems' security and stealing data. They are significantly transforming conventional phishing attempts into spear-phishing attacks, which often have disastrous outcomes for victims, in order to evade spam filters and detection techniques [11]. These targeted spear-phishing attacks are directed at specific individuals who meet certain criteria, including particular user groups, companies, or organizations. The content of the phishing emails is tailored to these targeted users, companies, or organizations, in contrast to spam emails and bulk phishing [11]. Unlike traditional phishing, attackers gather information about potential victims and conduct research to make it more likely that the emails they send contain convincing content [12].

## 1.5. Exercises and Drills in Cybersecurity

As mentioned in [13], drill and exercise are terms often used interchangeably, especially in the context of training sessions [14]. For the purpose of this study, "exercise" denotes methods utilized in practice, assessment, and training to enhance organizational performance. Additionally, exercises provide controlled opportunities to validate strategies, rules, and processes [15], aiding staff members in understanding their roles and responsibilities. According to [15], the National Institute of Standards and Technology (NIST) describes an exercise as a scenario-driven emergency situation simulation, designed to assess the efficacy of IT strategy components. Consequently, due to the varied interpretations of the term 'exercise,' cybersecurity exercises can take numerous forms, aiming to enhance or expand cybersecurity capabilities. These exercises can be conducted privately or publicly, often with the assistance of security experts. Ultimately, organizations benefit from tailored simulations to support cybersecurity within their operations [15].

As cited in [16], the term "cyberdrill" refers to a training process that simulates a cyberattack on workers or individuals whose occupations require responding to cybersecurity incidents [17]. It is described as simulating cybersecurity attack scenarios with a type of threat, and strategies to increase worker or user awareness of the risks [16]. Cyberdrills can also determine if a worker is particularly vulnerable to cybersecurity attacks. Rapid incident response enables the company to attain a particularly high degree of cybersecurity resilience against the effects of attacks. This response helps the business fulfill its commitments to customers, including internal clients, under the terms of the service level agreement (SLA). Cyberdrills can thereby increase employee awareness of cybersecurity issues and enable more effective responses. Workers should also receive regular training so they can make the best choices possible to lessen or completely eradicate the cyberthreats caused by phishing scams.

## 1.6. Previous Related Works

Numerous fascinating articles have been written about phishing, CSA, and the related variables. Table 1 presents the summarized information from the survey that is part of the literature review [5], [12], [17]-[33]. According to the literature review, although there are a few research papers conducted on the effects of different content or approaches sent with phishing attacks (e.g., [12], [24]), and there are a few research papers conducted on workers who work for critical information infrastructure (CII) entities (e.g., [12], [31]), there is still room to carry out a study to investigate this matter within a CII organization.

## 2. METHODOLOGY

This study was conducted within a Thai railway company primarily operating in the Bangkok area. This company is actually a state enterprise. Overall, there are approximately 800 staff, including employees and executives. For the purposes of this study, simulated cyber-attacks and cybersecurity awareness training were applied. The procedures in this study were broken down into the following three steps (see Figure 4):

1) Step 1: This step, which is called the first attack, involved creating the fictitious website depicted in Figure 5(a). It appeared to be one of the popular websites of a company that offers IT products (See Figure 5(b)). Following the creation of the fictitious website, 735 corporate workers (443 males and 292 females) received an email in Q2/2023 containing a fictitious promotion and a fictitious link. This step is also called the first simulation of a phishing attack associated with greed. Then, the results from the first attack were gathered.

2) Step 2: This step is called CSA training. The workers who fell victim to the initial phishing attack were provided with information on organizational and personal data protection during the CSA training described in this article. The training consisted of two sessions aboard the train, with each lesson lasting three hours. However, due to the COVID-19 situation in Thailand during that period, both training sessions were conducted online in Q3/2023, before conducting the third step.

Table 1. Related Works

| Authors & Reference | Country | Major issue |
|---|---|---|
| Daengsi et al. [5] | Thailand | The findings from this study demonstrated that Thai personnel working in technology and social-based departments within the same firm were aware of cybersecurity issues. The technology-based workers are the better than social-based workers, while the social-based workers were improved noticeably after they were involved with the CSA enhancement processes. |
| Abdullah and Mohd [12] | Malaysia | They used a spear phishing simulation to test the cybersecurity posture of companies in the telecommunication and defense subsector. The findings demonstrated that none of the samples fell for the spear phishing emails they received, suggesting that the general public had a high degree of cybersecurity awareness and knowledge. |
| Nachin et al. [17] | Thailand | They discovered that for reaching the CSA level in businesses, a simulation technique is more beneficial than an instructor approach. It is preferable to combine and utilize both options, though. |
| Cranor [18] | Unspecified | In this study, it was reported that about 12% of the victims associated with phishing clicked on bogus attachments and fake websites. |
| Chen et al. [19] | Unspecified | They presented that about 20% of a total of 250,000 workers in one well-known organization clicked on a malicious link. |
| Filippidis et al. [20] | Greek | They discovered that ISA was influenced by educational attainment and course of study. For instance, ISA was often higher among master's degree holders than among bachelor's degree holders. |
| Diaz et al. [21] | USA | They found a significant difference between the College of Natural and Mathematical Sciences and College of Arts, Humanities, and Social Sciences students' answers to the same phishing attack scenario. |
| Fatokun et al. [22] | Malaysia | Through their research, they found that age, gender, and educational background all have a big impact on students' cybersecurity habits. |
| Mousa [23] | Saudi Arabia | She found that Saudi Arabian students studying IT are not more talented than students studying other subjects. There was also found to be a significant difference between male and female students. |
| Li et al. [24] | USA | They discovered effects of phishing email content revealed significant effects on different age groups, on email types and marginally significant gender differences. |
| Greene et al. [25] | USA | According to research, it is important to train staff about phishing but also to investigate various approaches that would work best for each firm. |
| Carella et al. [26] | USA | They found that studying the impact of CSA training revealed that document training is superior to in-class instruction and no training. |
| Aljeaid et al. [27] | Saudi Arabia | They confirmed that users are vulnerable to attacks in the event that they lack cybersecurity knowledge and awareness, and they found that the type of attack technique affects users' views. |
| Davis and Grant [28] | USA | This study contrasted gamified phishing education games with phishing training exercises. The findings indicated that gamified phishing was more effective since it caught students' interest and improved their understanding. |
| Castaño et al. [29] | Unspecified | They created a program named PhiKitA to identify phishing and textual scams. The devised algorithm was 92.50% accurate, according to the results. |
| Sutter et al. [30] | Switzerland | In a 12-week study on phishing awareness training, more than 31,000 participants engaged in 144 distinct simulated phishing attempts. Data research indicates that 66% of users were not the intended victims of phishing attacks. |
| Rizzoni et al. [31] | Italy | They conducted a phishing simulation study with 6000 healthcare staff in a major Italian Hospital. The findings demonstrated that phishing emails with personalization are far more likely to be clicked on. |
| Hijji and Alam [32] | Pakistan | A framework for cybersecurity awareness and training was suggested by them. It is composed of twenty-five core routines and three main levels. Additionally, case studies are carried out, and the conclusions of the case studies suggest that the suggested framework can determine the skill levels of workers and assist in providing them with the training they need to successfully navigate cybersecurity concerns and challenges. |
| Hillman et al. [33] | Israel | According to the study in Israel with about 5,000 workers. They found that workers tend to engage with phishing emails that use personalized phrasing. Whereas the timing of training did not significantly affect phishing Click-Through Rate (CTR). Also, it is clear that training can enhance organizational CSA, while worker CSA and proactive behavior will continue to play a critical role. |

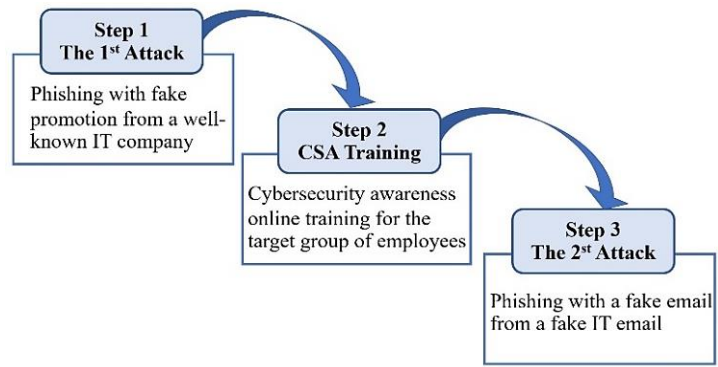

Figure 4. Overview on methodologies in this study

3) Step 3: This step is called the second attack, referred to as the second simulation of a phishing attack associated with fear. Utilizing different content from the initial attempt, this series of fear-mongering phishing attacks was conducted in Q3/2023. A fraudulent website was created to mimic the design of the organization's website. An email, tricking staff members into providing passwords, was distributed to every worker with a three-day deadline. Subsequently, the outcomes of the second attack were compiled.

After obtaining the data from both rounds of phishing simulations, the results from both simulations were presented in the next section. Additionally, a statistical test technique called Chi-square, which is a statistical calculation method used in examining the correlation between factors [35], was used to analyze the data from both simulations. The analyzed results were also presented in Section 3.

## 3. RESULTS, ANALYSIS AND DISCUSSION

As seen in Figure 6, this section displays the outcomes of two rounds of phishing attempts using descriptive statistics. In general, the outcome of the second strike is better than that of the first. The percentage of workers that fell into the trap decreased somewhat from 9.5% to 9.4% as shown in Figure 6(a) and 6(b) respectively, but both figures are still below the overall average of 12% and 20%, as stated in [18-19], respectively. It can be said that during the awareness training and, the proportion of workers within the entire organization that fell victim to phishing declined by approximately 1.05% (from (9.5-9.4-9.4)/9.5*100%).

After reviewing the response to emails of both attacks, it was found that just 1.4% of the workers had fallen victim to the two attacks. This indicates that the workers can learn new skills and become more aware through the knowledge transfer process. However, it was discovered that certain new workers had been caught in the second attack but not the first. This suggests that there are different ways in which phishing emails might harm staff members. For further investigation, the raw data were analyzed using the Chi-square technique, which is a form of analytical statistics, with the following hypotheses:

- $H1_0$: There is no significant relationship between gender and the first simulation of phishing attack.
- $H1_1$: There is a significant relationship between gender and the first simulation of phishing attack.
- $H2_0$: There is no significant relationship between gender and the second simulation of phishing attack.
- $H2_1$: There is no significant relationship between gender and the second simulation of phishing attack.
- $H3_0$: There is no significant relationship between two attack simulations and male workers.
- $H3_1$: There is significant relationship between two attack simulations and male workers.
- $H4_0$: There is no significant relationship between two attack simulations and female workers.
- $H4_1$: There is significant relationship between two attack simulations and female workers.
- $H5_0$: There is no significant relationship between two attack simulations and all workers.
- $H5_1$: There is significant relationship between two attack simulations and workers.

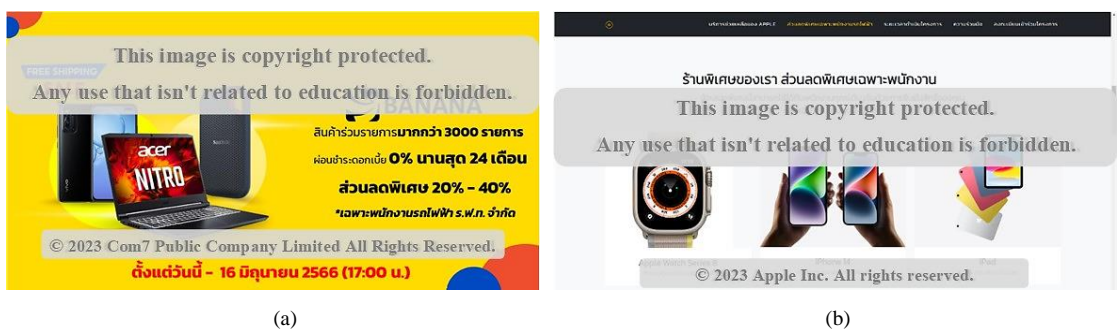

(a) (b)

Figure 5. The pages for the first attack (a) Fake promotion and (b) Products

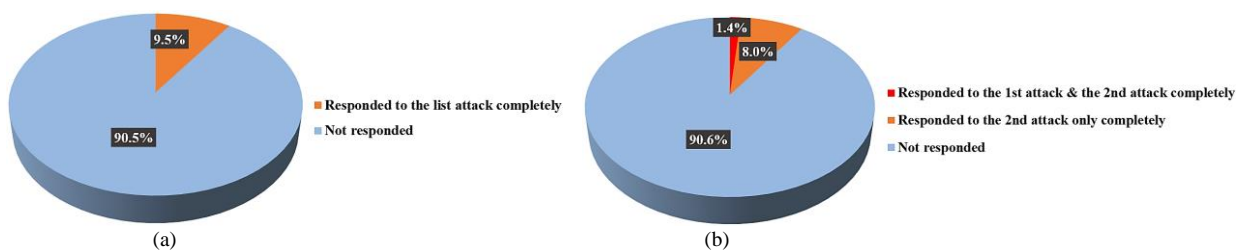

(a) (b)

Figure 6. The results from the phishing attack sumulation (a) The first attack (b) The the second attack

After conducting a statistical technique called Chi-square, the analyzed results from all hypotheses can be presented in Table 2. From the table, several issues (e.g., the relationship between the phishing attack and gender) can be discussed as follows:

Table 2. Analyzed Results from two Phising Attacks Using Chi-Square Technique

| Round of Attack | Gender | N | Chi-square | p-value | Remark |
|---|---|---|---|---|---|
| First | M vs F | 735 | 0.957 | 0.328 | Insignificant |
| Second | M vs F | 735 | 1.957 | 0.162 | Insignificant |
| First vs Second | M vs M | 443 | 38.025 | <0.001* | Significant |
| | F vs F | 292 | 260.004 | <0.001* | Significant |
| | All | 735 | 161.725 | <0.001* | Significant |

Note: 1) M and F stands for male and female respectively.
  2) * means it is significant with 95% confidence interval

• According to the hypothesis test result of H1, it was found that, for the relationship between the first simulation of phishing attack (associated with the content with greed) and gender, there is no significant difference (p-value = 0.328).

• For the hypothesis test result of H2, it was found that, for the relationship between the second simulation of phishing attack (associated with the content with fear) and gender, there is no significant difference (p-value = 0.162).

• For the hypothesis test result of H3, it was found that for the relationship between male workers and both phishing attack simulations, there is significant difference (p-value < 0.001).

• For the hypothesis test result of H4, it was found that there is significant difference (p-value < 0.001) for the relationship between female workers and both phishing attack simulations.

• For the hypothesis test result of H5, the last one, it was found that there is significant difference (p-value<0.001) for the relationship between all workers and both phishing attack simulations.

Last but not least, the study shows that regular cybersecurity training with different scenarios can improve employee knowledge or keep them up-to-date. However, these results are applicable only to a specific organization in Thailand; thus, they may not be representative of every organization associated with critical information infrastructure or CII. This means that they cannot be generalized to other CII organizations or other organizations in other sectors or countries. Thus, it is recommended that future work be done with other organizations or countries.

## 4. CONCLUSION

Due to cyberattacks, particularly phishing attacks, which are associated with cybersecurity and may seriously impact organizations, this study investigated the effects of phishing emails on railway workers in Thailand. It involved 735 workers (292 females and 443 males) from a CII railway firm. Interestingly, emails combining greed-inducing content in the first and fear-inducing content in the second had varying effects on the workers. Specifically, after analyzing the data using Chi-square, it was observed that gender differences did not play a significant role in each phishing attack. However, different content tricks did have a significant impact on phishing attacks. Therefore, it is crucial to conduct regular cybersecurity simulations and knowledge transfer using varied content to enhance workers' understanding of cybersecurity issues. This approach can help to improve the level of cybersecurity awareness among workers. Consequently, CII organizations will be better protected from cyberattacks, especially from phishing emails.

In future research, simulations incorporating additional tricks or types of content can be conducted for other critical information infrastructure organizations (e.g., banks and healthcare organizations) in Thailand. Additionally, organizations in other countries may also benefit from the methodology described in this study.

## ACKNOWLEDGEMENTS

Thank you to TRM-Plus Platform (grant number TMI2566-4026) for funding the ChaladOhn system development project. Furthermore, thanks to Rajamangala University of Technology Phra Nakhon, Rajamangala University of Technology Tawan-ok, and Rajamangala University of Technology Rattanakosin. Finally, thanks to Ms. Cecilia Mei-Yun Oh for English editing.

## BIOGRAPHIES OF AUTHORS

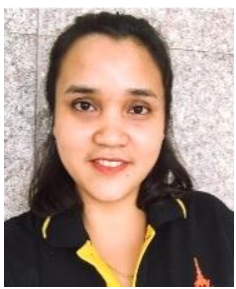

**Patsita Sirawongphatsara**    is a Lecturer in the Faculty of Science and Technology, Rajamangala University of Technology Tawan-ok (RMUTTO), Chonburi, Thailand. She received B.Sc. in computer science from RMUTTO in 2010 and M.Sc. in information technology at the King Mongkut's University of Technology North Bangkok (KMUTNB) in 2015. Her research interests include VoIP quality measurement, QoS/QoE, mobile networks, cybersecurity, data science, AI, and IoT. She is currently a Ph.D. student in the Faculty of Information Technology and Digital Innovation, KMUTNB. She can be contacted at email: patsita_si@rmutto.ac.th.

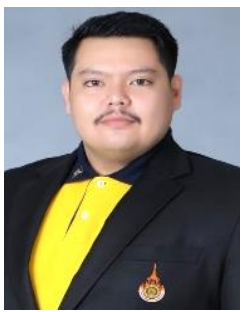

**Phisit Pornpongtechavanich**    is an Assistant Professor in the Faculty of Industry and Technology, Rajamangala University of Technology Rattanakosin Wang Klai Kangwon Campus (RMUTR_KKW), Thailand. In 2012, he received Bachelor of technology in information technology from RMUTR_KKW. He obtained a scholarship to study in Thailand and then received Master of Science in information technology from King Mongkut's University of Technology North Bangkok (KMUTNB), Thailand, in 2014. And he obtained a scholarship to study in Thailand and then received a Ph.D. in Information and Communication Technology for Education from KMUTNB in 2023. His research interests include security, deep learning, AI, VoIP quality measurement, QoE/QoS, mobile networks, and multimedia communication. He can be contacted at email: phisit.kha@rmutr.ac.th. (Noted: he is also the co-corresponding author for this paper.)

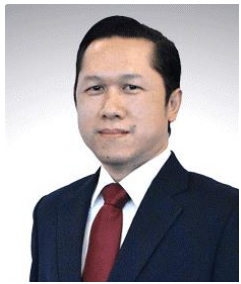

**Nattapong Phanthuna**    is an Assistant Professor in the Faculty of Engineering, Rajamangala University of Technology Phra Nakhon (RMUTP), Thailand. He received B.Eng. in Electrical Engineering from Rajamangala University of Technology in 1996 and M.B.A. in Industrial Management from Sripatum University, Thailand, in 1999. He received M.Eng. and D. Eng in Electrical Engineering from King Mongkut's Institute of Technology Ladkarbang (KMITL), Thailand, in 2007 and April 2011, respectively. At present, he is a Dean of the Faculty of Engineering, RMUTP. His main research interests include control system, illumination design, power electronic and image processing. He can be contacted at email: nattapong.p@rmutp.ac.th.

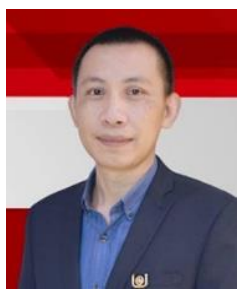

**Therdpong Daengsi**    is an Assistant Professor in the Faculty of Engineering, RMUTP. He received B.Eng. in Electrical Engineering from KMUTNB, Thailand, in 1997. He received a Mini-MBA Certificate in Business Management and M.Sc. in Information and Communication Technology from Assumption University, Thailand, in 2006 and 2008 respectively. Finally, he received Ph.D. in Information Technology from KMUTNB in 2012. He also obtained certificates including Avaya Certified Expert – IP Telephony and ISO27001. With 19 years of experience in the telecom business sector, he also worked as an independent academic for a short period before being a full-time lecturer at present. His research interests include VoIP, QoS/QoE, mobile networks, multimedia communication, cybersecurity, data science, and AI. He can be contacted at email: therdpong.d@rmutp.ac.th.